\documentclass{article}
\usepackage{graphicx}

\begin{document}

\title{\textbf{The Importance of Disagreeing}\\
Contrarians and Extremism in the CODA model}
\author{Andr\'e C. R. Martins and Cleber D. Kuba\\
Escola de Artes, Ci\^encias e Humanidades\\
Universidade de S\~ao Paulo -- Brazil}
\maketitle

\date{}
\begin{abstract}
In this paper, we study the effects of introducing contrarians in a model of Opinion Dynamics where the agents have internal continuous opinions, but exchange information only about a binary choice that is a function of their continuous opinion, the CODA model. We observe that the hung election scenario still exists here, but it is weaker and it shouldn't be expected in every election. Finally, we also show that the introduction of contrarians make the tendency towards extremism of the original model weaker, indicating that the existence of agents that prefer to disagree might be an important aspect and help society to diminish extremist opinions.
\end{abstract}

\section{Introduction}
\label{intro}

Contrarians~\cite{galam04,schneider04a,delalamaetal05a,zhongetal05a,borghesigalam06a,kurten08a,jacobsgalam08a} are agents who tend to disagree with other agentes, either locally or globally. The contrarian behavior is recognized as an investment strategy~\cite{corcosetal02a} when applied to Finance. While most Opinion Dynamics models~\cite{castellanoetal07} are based on the idea of agreement between the agents, the concept of contrarians was introduced in discrete models~\cite{galametal82,galammoscovici91,sznajd00,stauffer03a} by Galam~\cite{galam04} to help explain the hung election scenario, where elections seem to have a tendency to end much nearer to 50\% than it would be otherwise expected. 

Discrete models, however, are not perfectly suited to the description of extreme opinions, since they usually use only two opinions, as spins in the Ising model, with a few papers dealing with three possible choices~\cite{galam90a,galam91a,gekleetal05a}. If extreme opinions are to be incorporated, one option is to introduce inflexibles~\cite{galamjacobs07} in the system, that is, agents who don't change their minds under social influence. The effect of external social influence, when incorporated by the agents, can also be used as a source of extremism~\cite{galammoscovici91}.  Continuous models, like the Bounded Confidence models~\cite{deffuantetal00,hegselmannkrause02,lorenz07a}, on the other hand, are much better suited to this task, since it is easier to identify extreme values when the variable that measures opinions is continuous. This made the study of the spread of extremism in a society possible~\cite{weisbuchetal05,deffuant06}. 

In this paper, we will adopt a definition of extremism akin to that of Bounbded Confidence models, in the sense that we will use a variable $p_i$, that measures the probabilistic opinion of each agent. This opinion is a subjective probability associated with the proposition that one specific choice  in a binary situation is the best one. Therefore, it is easy to identify extreme beliefs as those were $p_i$ is very close either to 0 or to 1.

Bounded confidence models are also built above a continuous variable that can assume extreme values. Although useful in the description of the spreading of existing extremism, they are not capable of describing how extremism initially appears. A new class of models, using an external discrete observations that are actually a function of an internal, non-observable continuous opinion was proposed to avoid with this limitation. The Continuous Opinion and Discrete Actions model (CODA)~\cite{martins08a} used Bayesian rules~\cite{martins08e} to demonstrate a very simple update rule for the continuous variable and, as a result, the emergence of extremism was observed in an artificial society where no extremists existed at first. The model was further studied for different network structures and with the introduction of mobility and it was observed that extremist opinions are a pervasive characteristic of the model~\cite{martins08b}. A generalization of the model using neural networks was also used to investigate  the circumstances, in a cultural setting, where the model presented extremist factions and where it lead to consensus~\cite{vicenteetal08b}. 

In this paper, we will introduce heterogeneity in the agents of the CODA model, by allowing a proportion of those agents to be local contrarians. This will allow us to study the effects of the existence of contrarians in the overall extremism of the population. We will also investigate how the hung election scenario survives in the CODA model.

\section{Contrarians in the CODA model}
\label{sec:codacontrarians}

The CODA model is defined by a continuous variable $\nu_i$, $-\infty<\nu_i<+\infty$, that represents the continuous opinion of agent $i$ about one issue. The variable $nu_i$ is associated with the subjective probability $p_i$ that a given statement about the issue is true, according to agent $i$, by
\begin{equation}
\nu_i(t)=\ln\frac{p_i}{1-p_i}.
\end{equation}

When the agents interact, however, they do not observe each other's $\nu_i$ value, but the discrete choice associated with that value, $s_i= sgn(\nu_i)$. That is, while the opinions are continuous, the observed state of the system is expressed as a discrete Ising-like model.

Assuming symmetry between the two choices, the observation of the opinion of the neighbor $j$ by agent $i$ causes $i$ to change its internal value $\nu_i$ according to the simple rule
\begin{equation}\label{eq:dynamics}
\nu_i (t+1)= \left\{	
             \begin{array}{cc}
             			\nu_i (t)+a, & \textnormal{if } s_j=+1 \\
									\nu_i (t)-a, & \textnormal{if } s_j=-1,
									\end{array}
									\right.
\end{equation}
where $a$ is the size of the step for the dynamics of $\nu$. It should be noted that, as the step size is equal in both directions, the value of $a$ will have no influence at all at the dynamics of the $s_i$ variable. It does influence the probability value $p_i$ associated with $\nu_i$, but, since we will not measure that here, we will avoid assigning any value for $a$, keeping all measures of $\nu_i$ scaled as number of steps $\nu_i /a$ away from $\nu_i=0$.
Of course, the choice of agent $i$ will only be changed if the interaction causes $\nu_i$ to change sign. If not, the internal opinion might have become stronger or weaker (and more or less extreme), but the choice observed by the neighbors goes unchanged.

Contrarians can be easily introduced in this model by changing the additive term. If one multiply $a$ by a heterogeneity term $h_i$, such that $h_i =1$ stands for normal agents and $h_i=-1$ for contrarians, we will have the right effect. Notice that inflexibles can also be included by making $h_i=0$. It is easy to propose that we could have weak inflexibles, that is agents that change their opinion less. Those agents should just have a $h_i$ value in the interval $0\leq h_i<1$. Although this would make their values of $\nu_i$ (and, consequently, $p_i$) less extreme, the number of steps they would be from changing opinions, when we rescale their opinions by their own step size, would be unchanged for any non-zero value of $h_i$. Therefore, it makes sense to identify inflexibles only as individuals who don't change their minds at all.

\section{Simulations}
\label{sec:simulations}

\begin{figure}
 \resizebox{0.75\columnwidth}{!}{\includegraphics{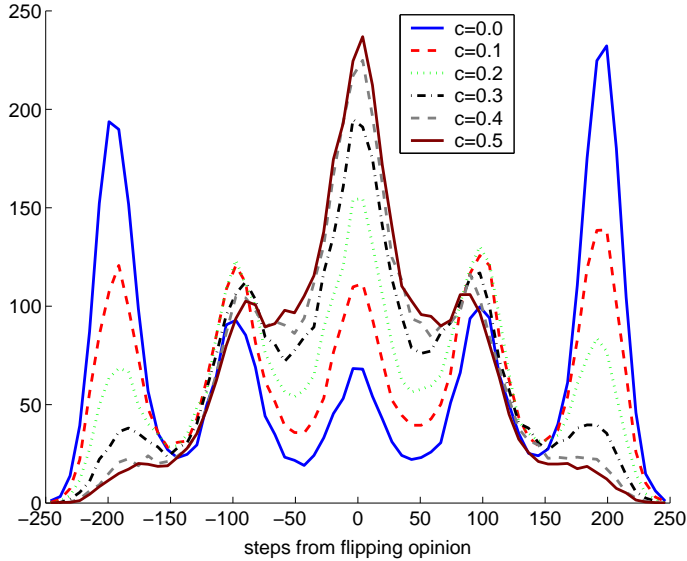}}
 \caption{Histogram for the average distribution of the agents continuous opinion $\nu_i$. Agents were located in a square 32x32 lattice with periodic boundary conditions. The histograms are shown for different proportions of contrarians, $c$, such that $0\leq c\leq 0.5$.}\label{fig:distrn64prop05voter}
 \end{figure}

\begin{figure}[ht]
 \resizebox{0.75\columnwidth}{!}{\includegraphics{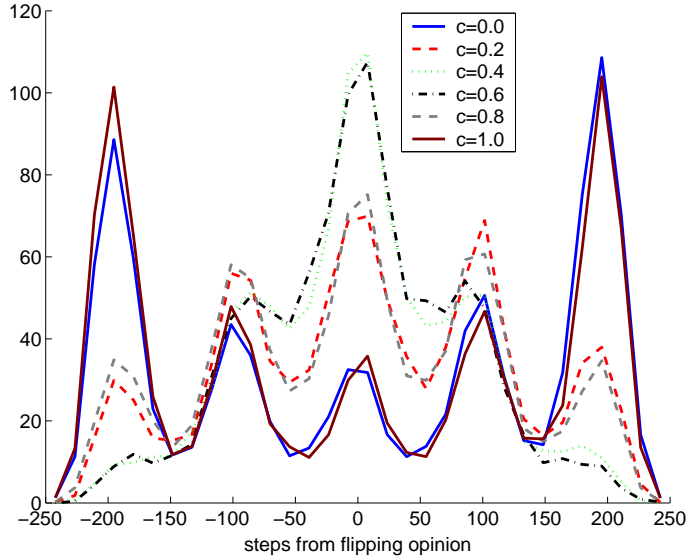}}
 \caption{Histogram for the average distribution of the agents continuous opinion $\nu_i$. Agents were located in a square 32x32 lattice with periodic boundary conditions. The histograms are shown for different proportions of contrarians, $c$, such that $0\leq c\leq 1.0$.}\label{fig:distrn64prop05voterallc}
 \end{figure}

In order to study the model, we have prepared a series of simulations for different proportions $c$ of contrarians. All realizations were in a bi-dimensional square lattice with periodic boundary conditions and a von Neumann neighborhood (four closest neighbors in a two dimensional square lattice), except when noted otherwise.
Figures~\ref{fig:distrn64prop05voter} and~\ref{fig:distrn64prop05voterallc} show the histograms for the final average distributions, where the continuous opinion $\nu_i$ was rescaled to show the number of steps the agent opinions are from $\nu_i=0$. The lattices used for those simulations had 32x32 agents each. 

In all the cases, initial conditions were such that each agent had an equal chance of supporting each choice ($s_i=+1$ or $s_i=-1$) and moderate values of $\nu_i$, that is $\nu_i<a$. That is, if in its first interaction, any agent were to observe one disagreeing neighbor, that agent would change their choices. Different curves correspond to different values of the proportion of contrarians $c$. Figure~\ref{fig:distrn64prop05voter} shows $c$ increasing with 0.1 steps, from 0.0 to 0.5, while, for Figure~\ref{fig:distrn64prop05voterallc}, $c$ increases in 0.2 steps, from 0.0 to 1.0.

One interesting feature of Figure~\ref{fig:distrn64prop05voter} is that, as $c$ increases up to 0.5, the extremist peaks become less important. At the same time, the central, moderate peak, becomes higher and much more important. Even for as few contrarians as $c=10\%$, this tendency is already observed, with the central peak becoming as high as the extremist peaks. The effect in favor of moderation keeps getting stronger until, for $c=50\%$, the extremist peaks are little more than a fat tail to a central distribution. Extreme opinions still survive, but the proportion of agents that have them becomes much smaller.
 
This could make one expects that, as $c$ becomes even larger, extremism might eventually disappear. A more complete range of values for $c$ (from $c=0.0$ to $c=1.0$) is shown in Figure~\ref{fig:distrn64prop05voterallc}. We can see clearly there that, as the proportion of contrarians increases even more, the effect is reversed and extremism is once more observed. We can also see a strong but unexpected symmetry for the histograms when one compares the histograms for $c$ and for $1-c$.

The observed symmetry is caused by the fact that, as more agents become contrarians and their proportion becomes too large, the lattice starts having a chess-like structure of choices, with agents alternating between $s_i=+1$ and $s_j=-1$. When that happens, the contrarians reinforce their opposite opinions, the same way that normal, agreeing agents would reinforce their opinions inside a cluster of equally minded individuals.

 \begin{figure}[ht]
 \resizebox{0.75\columnwidth}{!}{\includegraphics{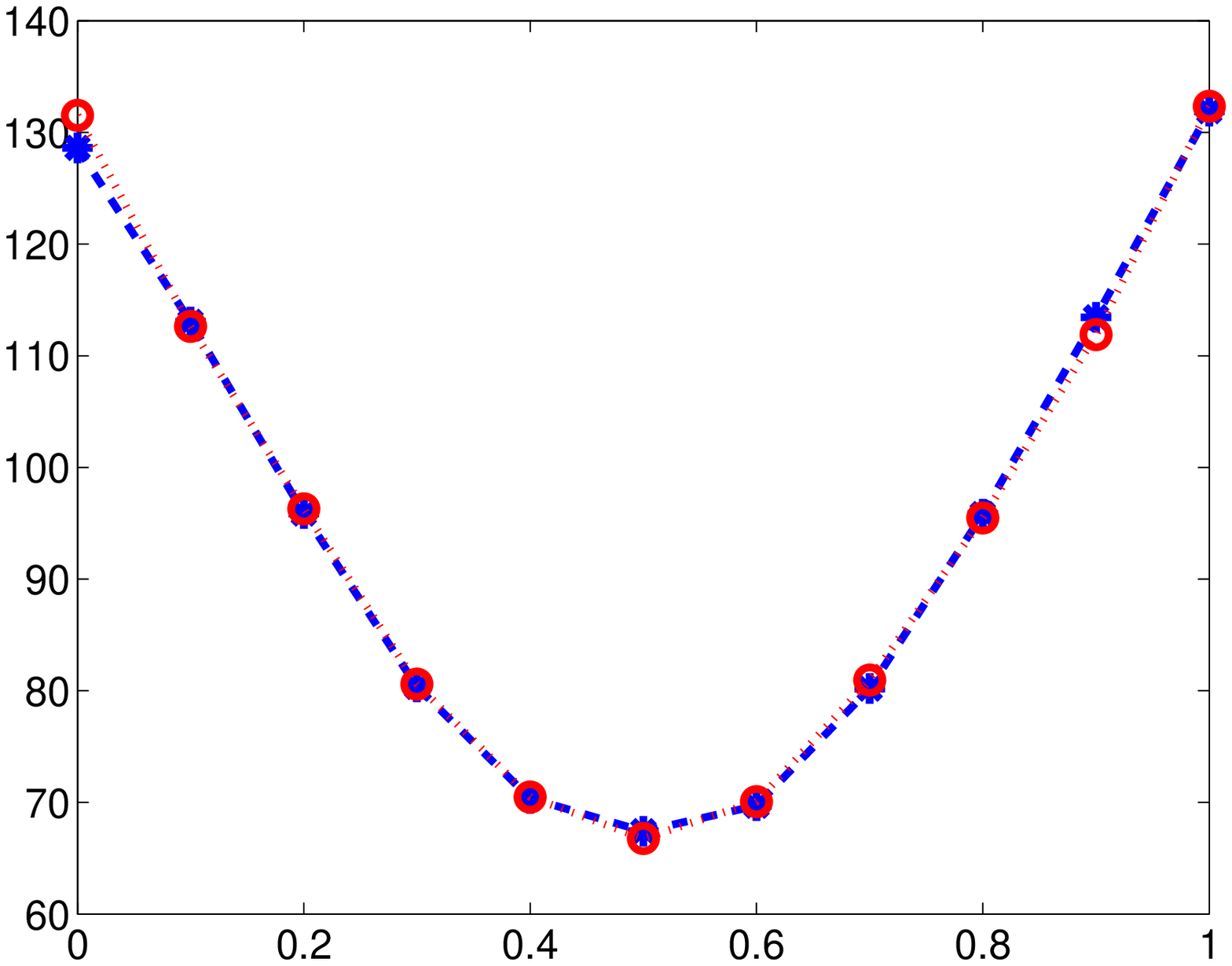}}
 \caption{Spread of the final continuous opinions, measured by the standard deviation of the opinions, as a function of the percentage of contrarians, $c$ for both 32x32 (blue curve) and 64x64 lattices (red)}.\label{fig:contrspread3264voter}
 \end{figure}

We can also measure the importance of extremism in the final configuration of the system by calculating the standard deviation of all $\nu_i$ values. Those standard deviations are shown in Figure~\ref{fig:contrspread3264voter} as a function of $c$, for both 32x32 (blue curve) and 64x64 lattices (red). Notice that both curves present a remarkable agreement, indicating that the number of agents has no influence or a very weak one on the behavior of the system. We can also clearly see the decrease in the spread of the opinions, with minimum extremism observed around $c=0.5$. The symmetry between $c$ and $1-c$ is also clear. However, one should notice that, while the standard deviation does become smaller, it only decreases to around half its highest value. This corresponds to our previous observation that, while extremism does decrease, it is still there and the proportion of extremists in the tails can't be ignored.

\begin{figure}[ht]
 \resizebox{0.75\columnwidth}{!}{\includegraphics{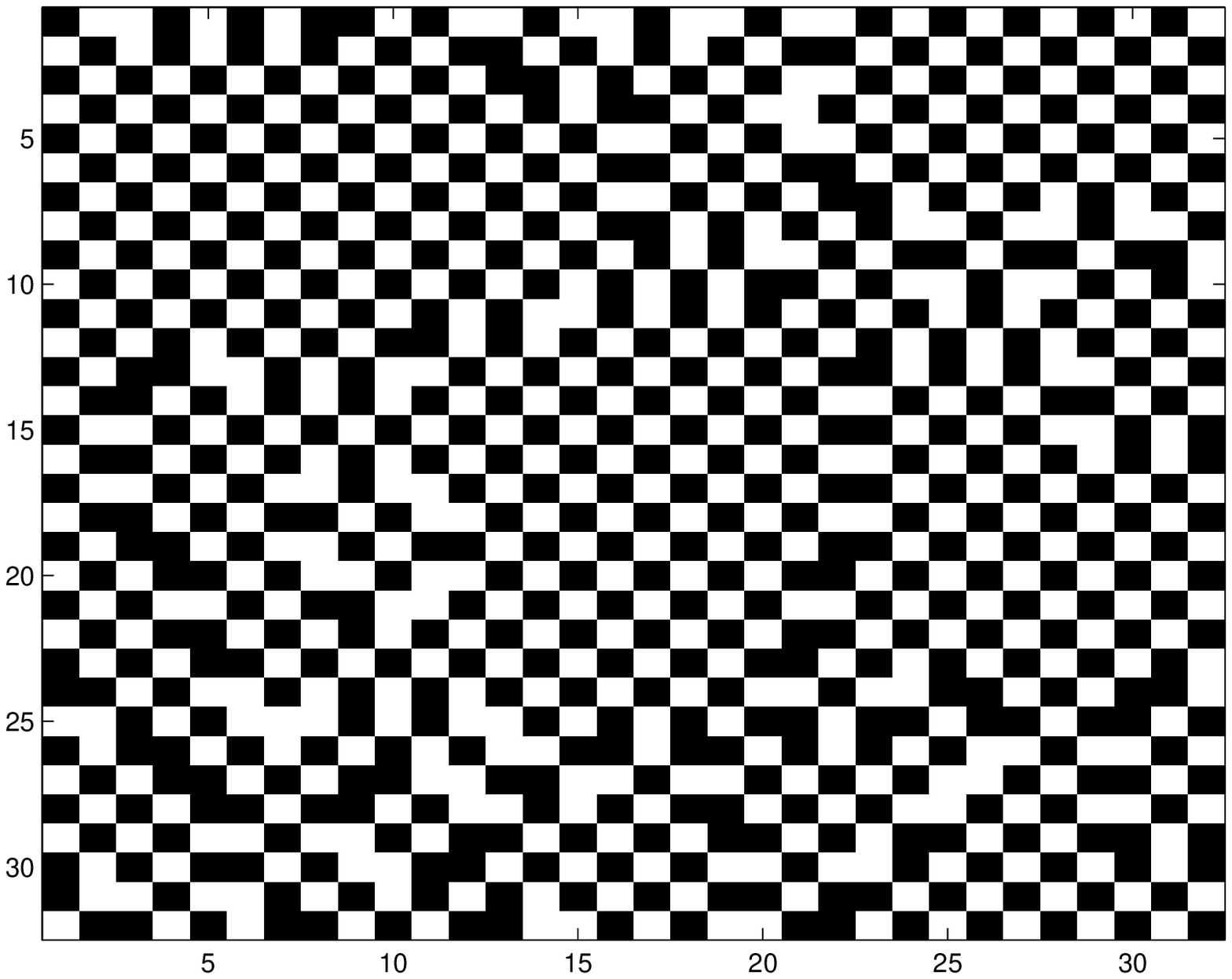}}
 \caption{Final configuration for the regular square lattice with 32x32 agents, with a von Neumann neighborhood and $c=100\%$ contrarians. }\label{fig:latpcont10votern32}
 \end{figure}
 
As mentioned before, the effects observed for large proportion of contrarians ($c>50\%$) are due to the system finding a stable spatial structure. Figure~\ref{fig:latpcont10votern32} shows that structure when $c=1.0$. It shows clearly that the chessboard structure is the equivalent to the consensus in the $c=0.0$ case. Notice that, for a chessboard, every contrarian is surrounded only by agents who disagree with him. That always reinforces the contrarian opinion and the contrarian eventually becomes a extremist. The breaks in the structure are equivalent to the interfaces between the opinion domains.
 
\begin{figure}[ht]
 \resizebox{0.75\columnwidth}{!}{\includegraphics{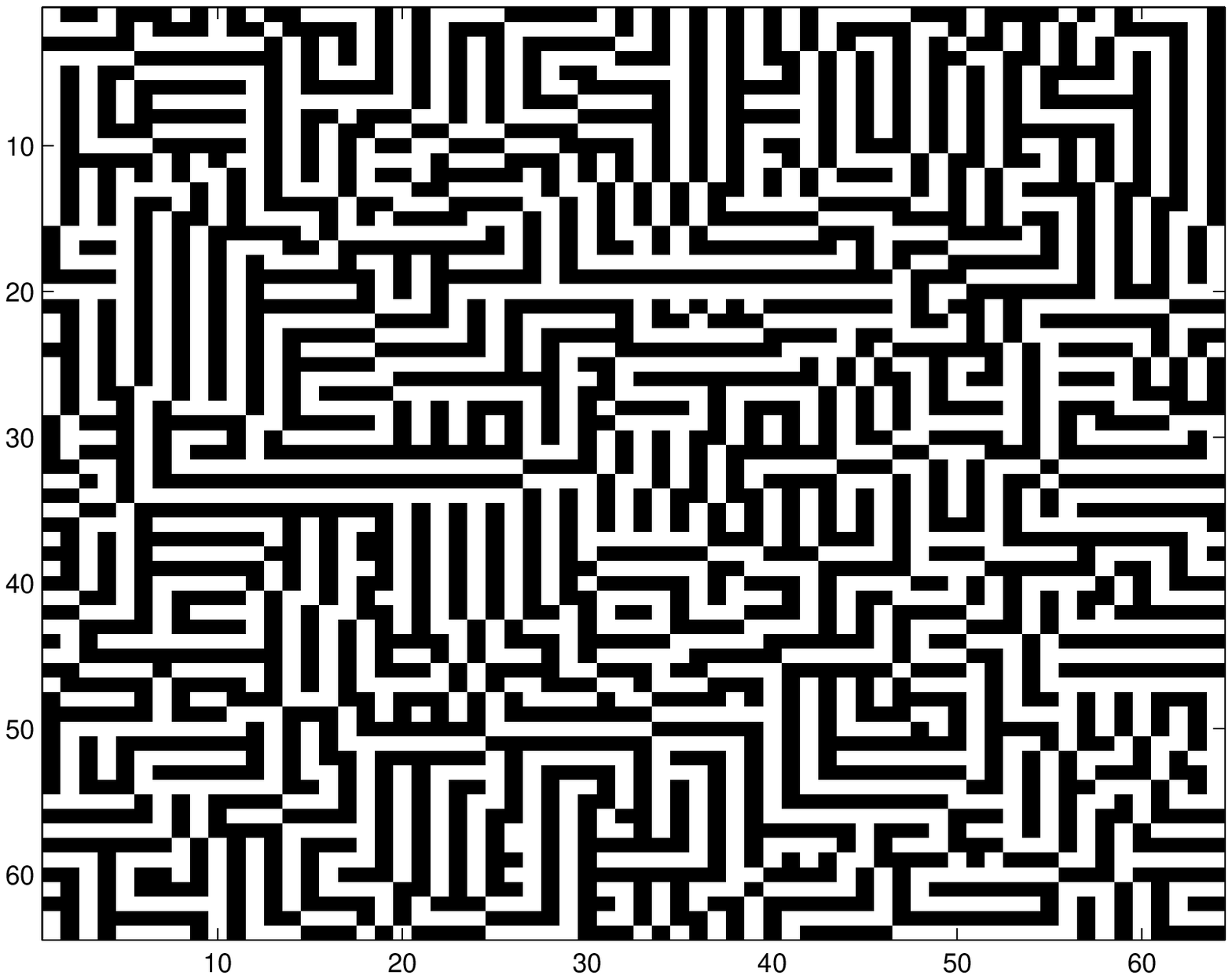}}
\caption{Final configuration for the regular square lattice with 32x32 agents, with a Moore neighborhood and $c=100\%$ contrarians. }\label{fig:n64d2finalconf}
 \end{figure}

This suggests that this final observed configuration might depend heavily on the chosen social structure that was used. Different social networks could prevent stable configurations from forming and that could destroy the observed symmetry for large values of $c$. In order to test this possibility, the simulations were repeated for a Moore neighborhood (including the diagonal neighbors, for a total of 8 neighbors in a bi-dimensional square lattice). This certainly destroys the reinforcement of opinions in Figure~\ref{fig:latpcont10votern32}, since the chessboard structure would mean four disagreeing neighbors, but also four agreeing ones.

The final configuration for the Moore network can be seen in Figure~\ref{fig:n64d2finalconf}. It is interesting to notice that the system finds another stable state where most agents have their opinions reinforced. No neighborhood consisting only of disagreeing neighbors is possible any more. However, parallel lines with alternating opinions mean each agent has 6 disagreeing neighbors and only 2 neighbors that think the same. This obviously mean that the reinforcement will not be as strong as when $c=0$, since domains of equally minded agents have all neighbors agreeing. That shows that the symmetry between $c$ and $1-c$ observed in Figure~\ref{fig:distrn64prop05voterallc} was just an artifact of the network and, with a different network, it is now destroyed. But since the system finds a stable state where some reinforcement of opinions still happen, we should expect that some extremism should be observed in this case.

The introduction of contrarians has also an important effect in the CODA model. Without them, the system has fixed points where all agents make the same choice $s_i$ and complete consensus emerges. The system was usually prevented from reaching that state due to the strengthening of the domain walls, but full consensus was always a trivial solution to the problem. With contrarians, consensus is no longer a fixed point. Even if it were reached somehow, the contrarians would soon be influenced into adopting the opposite point of view. This suggests that it is also interesting to study the effects that contrarians have in the proportion of agents that adopt the majoritary position. 

\begin{figure}[ht]
 \resizebox{0.75\columnwidth}{!}{\includegraphics{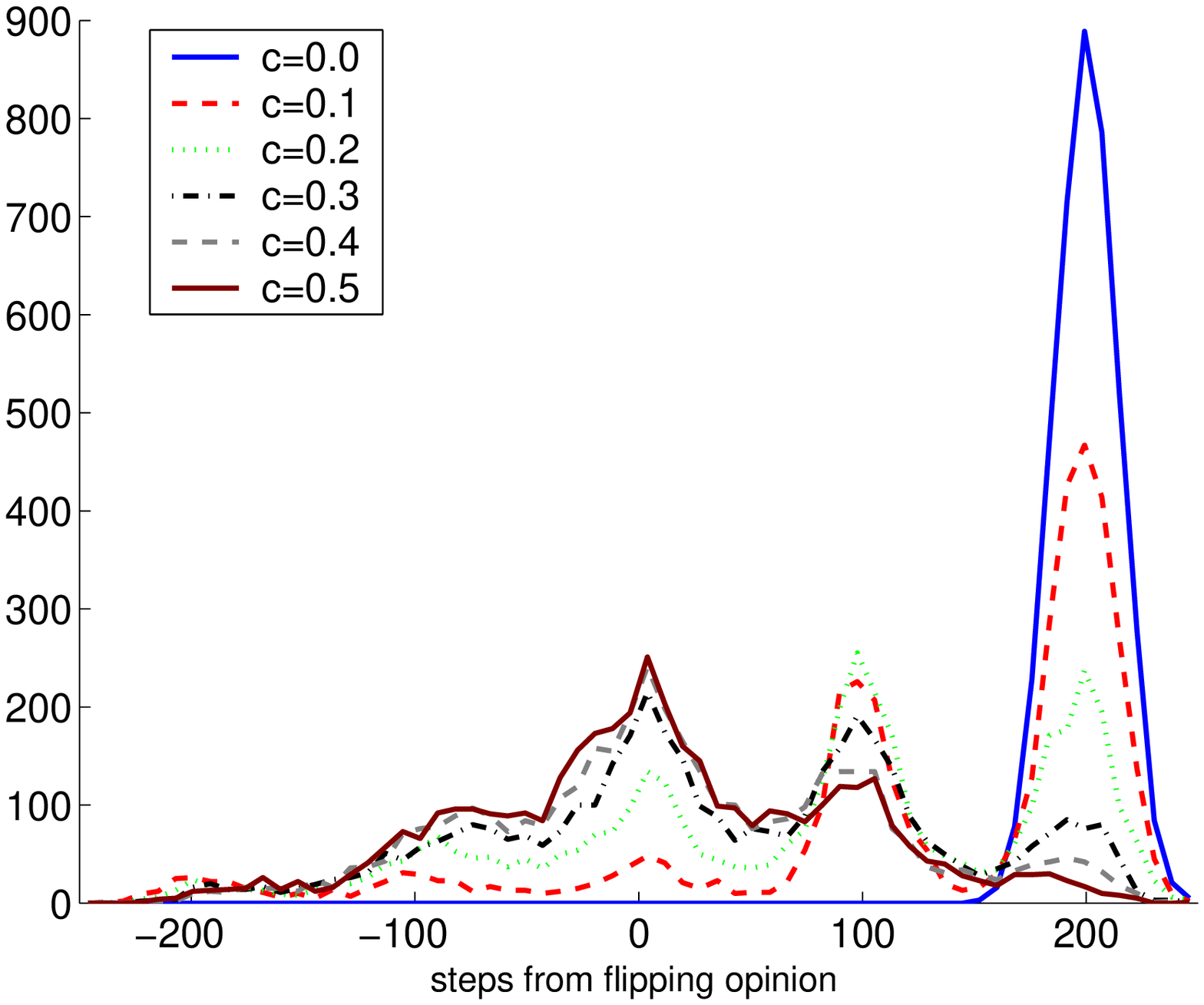}}
 \caption{Histogram with the average distribution of agent opinions. Agents are distributed in a 32x32 square lattice and, as initial condition, 95\% of the agents supported the same alternative. Different curves correspond to different values of $c$, so that $0\leq c\leq 0.5$.}\label{fig:distrn64prop095voter}
 \end{figure}

For discrete models, it is already known that above a certain proportion, we observe the hung election scenario, where the system as a whole tends to a tie~\cite{galam04}. Figure~\ref{fig:distrn64prop095voter} shows simulations where the initial conditions were no longer symmetric. Instead, each agent had a 95\% chance of choosing $s_i=+1$. All agents were moderate at first and curves for values of $c$ between 0 and 0.5 are shown.

Some new features can easily be observed. First, for $c=0$, the majority is large enough to ensure that consensus emerges and one no longer has two opposing factions as final state. As we discussed, as soon as some contrarians are introduced, the possibility of observing consensus is destroyed and we always have the survival of a minority of extremist agents who oppose the majority. Also, a second peak appears between the moderate and the extremist majority peak appears. This second peak correspond to agents who have, among their neighbors, three agents who make them want to support the majority and one that influences them the other way. That is, it is also an artifact of the network.  

 \begin{figure}[ht]
 \resizebox{0.75\columnwidth}{!}{\includegraphics{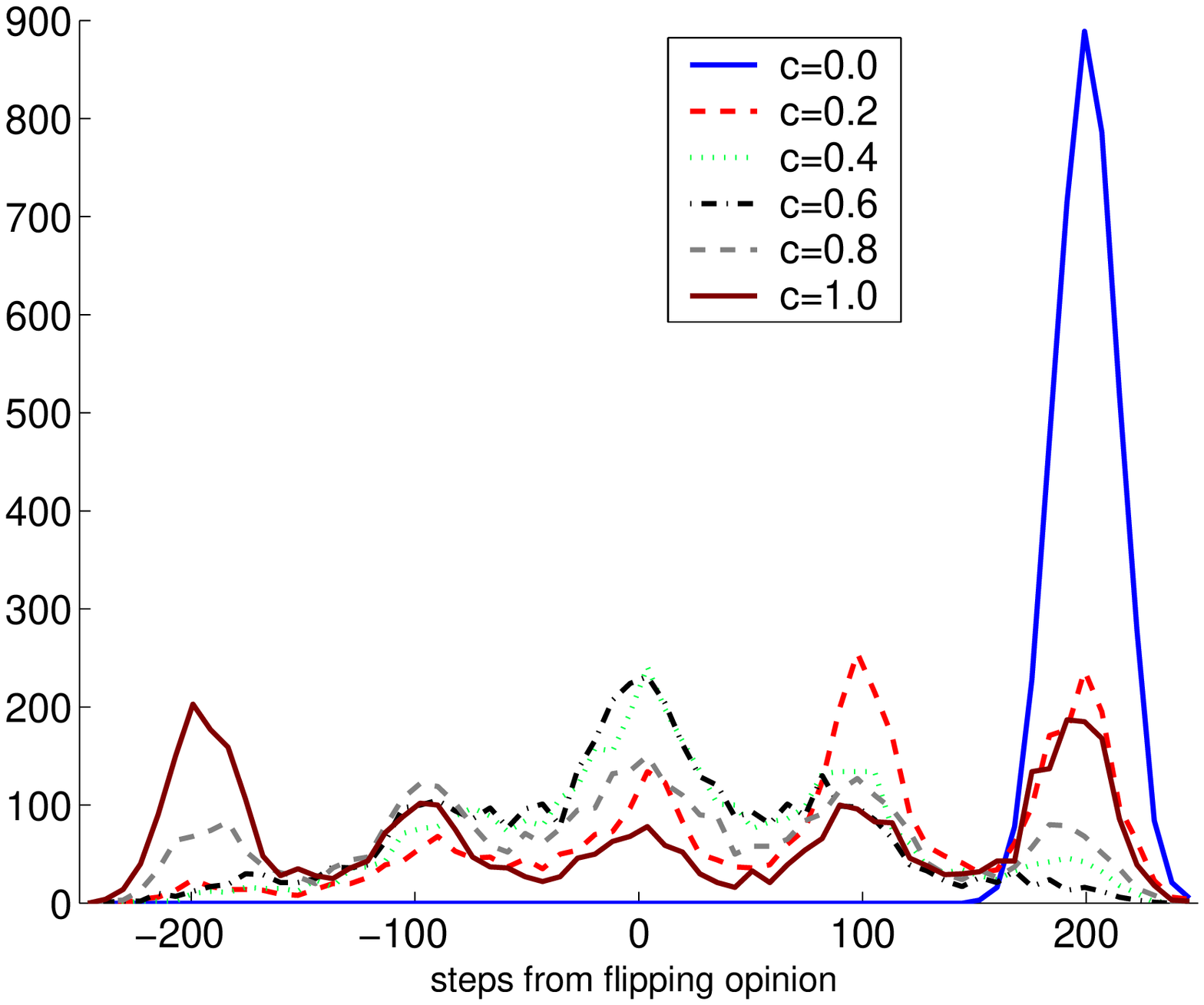}}
 \caption{Histogram with the average distribution of agent opinions. Agents are distributed in a 32x32 square lattice, with a von Neumann neighborhood and, as initial condition, 95\% of the agents supported the same alternative. Different curves correspond to different values of $c$, so that $0\leq c\leq 1.0$.}\label{fig:distrn64prop095voterallc}
 \end{figure}

As $c$ keeps increasing, the extremist minority peak does not increase, but starts decreasing. As we get to $c=0.5$, the central distribution we had observed in Figure~\ref{fig:distrn64prop05voter} is recovered. We still have the survival of the initial majority peak as well as of the second intermediary peak, but both them lose their initial strength.

In order to investigate possible effects similar to the symmetry of Figure~\ref{fig:distrn64prop05voterallc}, we can see the results for the same problem (95\% of initial majority) in Figure~\ref{fig:distrn64prop095voterallc}, for values of $c$ between 0.0 and 1.0. The symmetry between $c$ and $c$ is indeed destroyed and we can watch that, for values of $c$ above 0.5, we basically recover the results observed for initial conditions with half of the agents supporting each choice. That is, the effect of the initial conditions seems to be destroyed by the large number of contrarians.

\begin{figure}[ht]
 \resizebox{0.75\columnwidth}{!}{\includegraphics{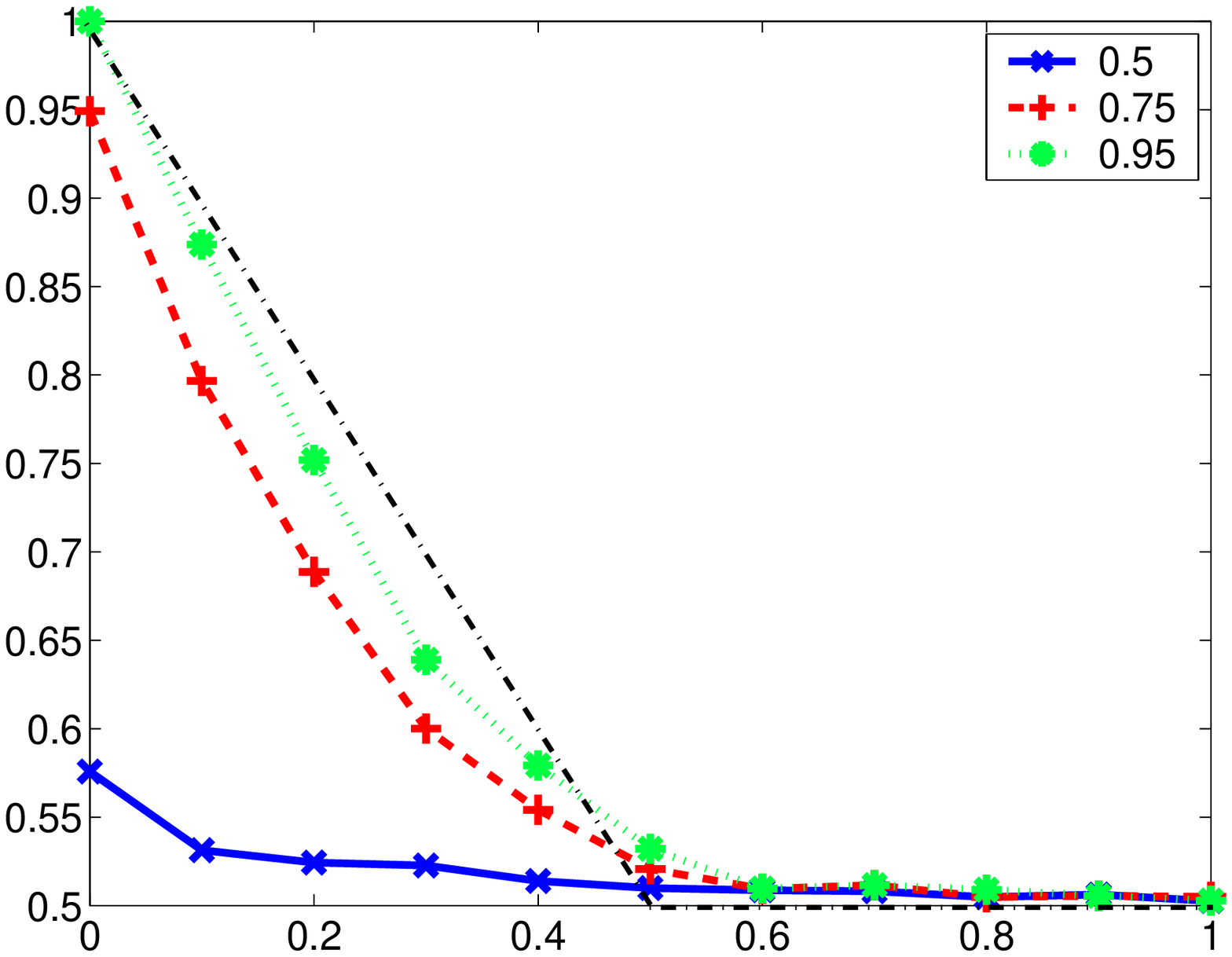}}
 \caption{Final proportion of agents that agree with the observed majority, as a function of $c$. Different curves correspond to different initial proportion of agents that support the majority. The black line with dashes and dots correspond to the syncronous mean field solution presented at Section~\ref{sec:meanfield}}\label{fig:propmajn32voter}
 \end{figure}

This is similar to the hung election scenario~\cite{galam04}, where, above a certain threshold, the existence of contrarians always led the system to the fixed point where half of the agents supported each choice. And we have, indeed, observed the tendency towards 50\% in our results so far, even for a large initial proportion of 95\% of the agents supporting the majority. This tendency seems robust, but we still need to investigate it. Figure~\ref{fig:propmajn32voter} shows the final proportion of agents who agree with the majority as a function of $c$, for three different initial proportions of majority supporters,  50\%, 75\% and 95\%. 

We observe that, for values of $c$ larger than 0.5, all scenarios show final proportions very close to a tie, as in the hung election scenario. For $c$ smaller than 0.5, the tendency towards a tie is still observed, with the diminishing size of the majority, but no hung elections are really observed. It is also interesting to notice that, even for initial conditions starting at a tie, more contrarians make the final state closer to 50\% than before. That is, their introduction makes even the natural statistical variance around the expected value smaller. A few extra simulations were performed for a larger number of agents and, for the cases close to 50\%, the difference of the actual 50\% value was smaller, indicating that the small differences we observe in Figure~\ref{fig:propmajn32voter} are due to statistical error.

\begin{figure}[ht]
 \resizebox{0.75\columnwidth}{!}{\includegraphics{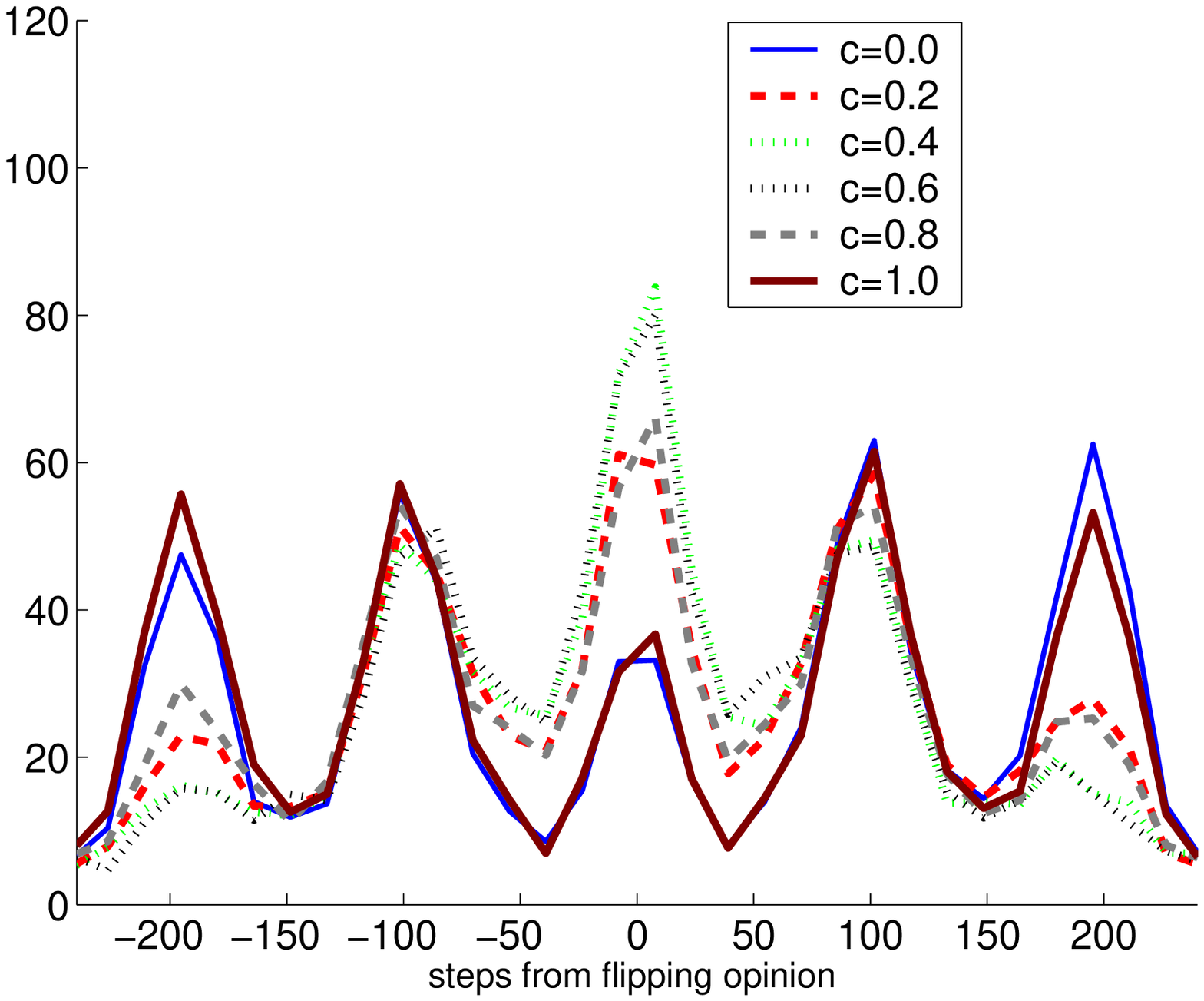}}
 \caption{Histogram of the average distribution of agent opinions. Agents are distributed in a 32x32 square lattice, with a von Neumann neighborhood and the different curves correspond to different values of $c$, so that $0\leq c\leq 1.0$. Here, 30\% of the agents are inflexibles and do not change their opinions at all ($h_i=0$). Initially, each choice was supported by half of the agents.}
 \label{fig:distrn32infl03}
 \end{figure}

Finally, since it is very simple to implement, we have also explored the case where inflexibles were introduced in the system. Figure~\ref{fig:distrn32infl03} shows an initially symmetric case where 30\% of the agents did not update their opinions. Since those agents do not change, the figure shows the observed distribution just for the non-inflexibles. These results should be compared with the similar case, with no inflexibles, shown in Figure~\ref{fig:distrn64prop095voterallc}. It is easy to notice a similar shape of the curves, with some diminishing of the extremism arising from the fact that there are many agents who won't change their minds. Often, some of those will end up in the middle of a domain of opposing view, where they will make the local choice weaker. Larger proportions of inflexibles would mean that very few agents update their minds, making it a less interesting case. A few simulations were run, showing no new interesting results for the inflexibles, other than this diminishing of the extremist positions.

\section{Mean Field solution}\label{sec:meanfield}

So far, the dynamics of the simulated systems was asynchronous, with one agent updating its mind at each time. However, we can also obtain a mean field result trivially. If one starts with a proportion of $r$ agents that support the majority (contrarians or not) and a proportion $p$ of agents who are contrarians, we only have to observe the mean field effects on each type of agent, that is a typical contrarian and a typical normal agent. Before the first interaction step, assuming initial independence between being a contrarian and the support for the majority,  $rc$ proportion of contrarians in the majority, $(1-r)c$ contrarians in the minority, $r(1-c)$ non-contrarians in the majority, and $(1-r)(1-c)$ non-contrarians in the minority.

Since the average effect is felt by each agent, the contrarians in the minority as well as the non-contrarians in the majority will have their opinions reinforced. On the other hand, the contrarians in the majority will start shifting their internal opinion towards the minority and the non-contrarians in the minority will also change, but towards majority. Assuming everyone starts at the same distance from changing opinions, the agents moving towards the opposite view will change opinion at the same time, since the mean field has equal value for everyone (only the direction of the change is reversed). At the end, the system stabilizes with contrarians in the minority and the non-contrarians in the majority and we have a majority with a proportion of $1-c$, regardless of $r$. 

One should notice that this argument is only true as long as $c\leq 0.5$. If the majority of the agents are contrarians, when the contrarians in the minority opinion change their opinions, all contrarians will be in the majority and all non-contrarians will be in the minority. Therefore, in the next step, all contrarians will reverse the process. The system will actually oscillate around 50\%, with no friction force to make it stop. This tendency of returning to 50\% makes it reasonable to conclude that the time average will be around 50\%.

Figure~\ref{fig:propmajn32voter} showed a black line, with dots and dashes, with the solution that the final proportion in the majority is $1-c$ for $c\leq 50\%$ and $50\%$ for $c>0.5$. We can see that the simulated results show a similar but not identical behavior. The local effects and the asynchronous character of the simulation, characteristics absent from the mean field result, seem to make the final proportion of the majority smaller than it would be otherwise.

\section{Conclusion}\label{sec:conclusion}

Here we have studied the effects of introducing contrarians in the CODA model. We have seen that the contrarians have an important effect in diminishing extremism, as long as they are not the majority, when they might create a different kind of extremism where everyone wants to disagree with their neighbors. It is reasonable however to assume that contrarians are not so common and, under that assumption, an increase in the proportion of people willing to disagree might be beneficial in the cases where it would be better if there were less extremism.

This effect does come with a price, since contrarians prevent consensus from ever being reached in a society. However, a good case can be made that democracy actually needs dissent~\cite{huckfeldtetal0} in order to avoid totalitarianism and, therefore, this consensus preventing effect can also be seen as a positive one, when the debate is about political issues.

Finally, we have analyzed how the hung election scenario survives in the CODA model. While the hung elections are still expected for a majority of contrarians, the effect is not so strong as that observed in discrete models for more reasonable amounts of contrarians. We should point out that the contrarians did decrease the tendency to consensus and made the final results closer to 50\%, but not exactly 50\%. This seems to suggest that, while hung elections should be more common than statistically expected, not every election should present a result as strong as that predicted by the discrete models. And, as a matter of fact, it is true that humg elections seem common, but not unavoidable.

\section{Acknowledgement}

One of the authors (ACRM) would like to thank Funda\c{c}\~ao de Amparo \`a  Pesquisa do Estado de S\~aoPaulo (FAPESP) for the support to this work, under grant 2008/00383-9.


 \bibliographystyle{epj}
 \bibliography{biblio}

\end{document}